\documentclass[sigconf]{acmart}

\AtBeginDocument{%
  }

\usepackage{optidef}
\usepackage{enumerate}
\usepackage{subcaption}
\usepackage{enumitem}

\copyrightyear{2025}
\acmYear{2025}
\setcopyright{acmlicensed}\acmConference[CIKM '25]{Proceedings of the 34th ACM International Conference on Information and Knowledge Management}{November 10--14, 2025}{Seoul, Republic of Korea}
\acmBooktitle{Proceedings of the 34th ACM International Conference on Information and Knowledge Management (CIKM '25), November 10--14, 2025, Seoul, Republic of Korea}
\acmDOI{10.1145/3746252.3761567}
\acmISBN{979-8-4007-2040-6/2025/11}

\begin{document}

\title{Learning to Comparison-Shop}

\settopmatter{authorsperrow=3}
\author[Jie Tang]{Jie Tang}
\orcid{0009-0006-9509-2367}
\email{jie.tang@airbnb.com}
\affiliation{%
  \institution{Airbnb}
  \city{San Francisco}
  \country{USA}
}

\author{Daochen Zha}
\orcid{0000-0002-6677-7504}
\email{daochen.zha@airbnb.com}
\affiliation{%
  \institution{Airbnb}
  \city{San Francisco}
  \country{USA}
}

\author{Xin Liu}
\orcid{0009-0008-7563-0425}
\email{xin.liu@airbnb.com}
\affiliation{%
  \institution{Airbnb}
  \city{San Francisco}
  \country{USA}
}

\author{Huiji Gao}
\orcid{0009-0006-0424-248X}
\email{huiji.gao@airbnb.com}
\affiliation{%
  \institution{Airbnb}
  \city{San Francisco}
  \country{USA}
}

\author{Liwei He}
\orcid{0009-0007-2942-3145}
\email{liwei.he@airbnb.com}
\affiliation{%
  \institution{Airbnb}
  \city{San Francisco}
  \country{USA}
}

\author{Stephanie Moyerman}
\orcid{0000-0003-0526-1077}
\email{stephanie.moyerman@airbnb.com}
\affiliation{%
  \institution{Airbnb}
  \city{San Francisco}
  \country{USA}
}

\author{Sanjeev Katariya}
\orcid{0009-0008-1519-174X}
\email{sanjeev.katariya@airbnb.com}
\affiliation{%
  \institution{Airbnb}
  \city{San Francisco}
  \country{USA}
}

\renewcommand{\shortauthors}{Jie Tang et al.}

\begin{abstract}
 In online marketplaces like Airbnb, users frequently engage in comparison shopping before making purchase decisions. Despite the prevalence of this behavior, a significant disconnect persists between mainstream e-commerce search engines and users’ comparison needs. Traditional ranking models often evaluate items in isolation, disregarding the context in which users compare multiple items on a search results page. While recent advances in deep learning have sought to improve ranking accuracy, diversity, and fairness by encoding listwise context \cite{Weiwen22}, the challenge of aligning search rankings with user comparison shopping behavior remains inadequately addressed. In this paper, we propose a novel ranking architecture — Learning-to-Comparison-Shop (LTCS) System — that explicitly models and learns users’ comparison shopping behaviors. Through extensive offline and online experiments, we demonstrate that our approach yields statistically significant gains in key business metrics — improving NDCG by 1.7\% and boosting booking conversion rate by 0.6\% in A/B testing — while also enhancing user experience. We also compare our model against state-of-the-art approaches and demonstrate that LTCS significantly outperforms them. 
\end{abstract}



\maketitle

\section{Introduction} \label{intro}
 Online marketplaces such as Amazon, eBay, Walmart, and Airbnb provide search functions that help users discover products or services tailored to their preferences. Despite offering different services, from retail goods to food delivery and rental accommodation, these platforms share a common goal: their ranking models are designed to present the most relevant items to users. However, from the user’s perspective, the need for comparison shopping often persists before arriving at a final decision. Comparison shopping helps users achieve an optimal balance of price, quality, and convenience when making purchase decisions.\par
 Despite users’ need for comparison shopping, mainstream ranking models are typically designed without accounting for this behavior. Traditional learning-to-rank models are primarily built to learn a scoring function that computes a relevance score for each item independently. In other words, these models lack the ability to directly compare items with one another during inference, even though such comparisons are often incorporated into offline training through pairwise \cite{Burges05} \cite{burges2010} or listwise loss functions \cite{cao07}. Recent efforts \cite{Rama20} \cite{Qingyao19} \cite{Liang20} have introduced attention mechanisms to capture cross-item interactions in learning-to-rank models. These approaches encode ranking context, enabling the model to directly compare items during inference. However, merely encoding ranking context is insufficient to accurately capture and learn users’ comparison shopping behavior. \par
 To better understand this behavior, we can take a step back and consider how users engage in comparison shopping: initially, users browse the top results returned by a search ranking model, evaluating each item independently to determine whether to click on it (a pointwise approach). After selecting items, users compare the clicked items by examining various details, including descriptions, prices, reviews, images, and more. This process often involves several rounds of back-and-forth deliberation, which may span minutes, hours or days before users reach a final decision. \par
 Encoding ranking context can certainly enhance the comparison stage. In a traditional multi-stage ranking system, one might train a ranking model (initial ranker) to learn users' pointwise clicking behavior, while a separate re-ranker \cite{Weiwen22} employs ranking context encoding to learn users' comparison behavior. However, this approach is fundamentally distinct from actual user comparison shopping behavior. From the user's perspective, pointwise clicking and comparison behavior are highly interconnected, with the rationale behind pointwise clicking significantly influencing the comparison process. Training the ranker and re-ranker separately risks disrupting this connection; in some instances, they may even conflict with one another.\par
 In this paper, we propose a novel model that efficiently learns users' comparison shopping behavior. To the best of our knowledge, this work represents the first attempt to address the comparison shopping problem within the learning-to-rank domain. Our major contributions include:
  \begin{itemize}
      \item We decompose users’ comparison shopping behavior into two distinct stages: a pointwise evaluation stage and a set-wise comparison stage.
      \item We propose a novel framework that models and learns users’ two-stage comparison shopping behavior as an integrated process.
      \item Behavior-aligned co-training: We propose a novel co-trained architecture that mirrors users’ two-stage comparison behavior by combining a pointwise initial ranker with a setwise re-ranker.
      \item We successfully applied this framework to a large-scale system, such as Airbnb’s search ranking system. Additionally, we conducted thorough analyses to validate our design assumptions and choices.
  \end{itemize}

While both setwise reranker and co-training approaches have been individually explored in ranking models, our key contribution lies in their behavior-aligned integration to simulate the two-stage comparison shopping process common in online marketplaces. Specifically, our LTCS framework jointly trains a lightweight, pointwise initial ranker and a transformer–based setwise re-ranker to reflect how users first evaluate listings independently, then compare top candidates before making a decision. This co-trained architecture not only improves ranking performance but also enhances consistency between initial and final ranking stages—something prior work such as SetRank and other neural re-rankers do not address, as they typically model listwise interactions without an intermediate filtering stage or coordinated training. To our knowledge, LTCS is the first LTR architecture that explicitly models and learns from this comparison-driven user behavior, leading to both significant offline gains and measurable business impact in production.

 \begin{figure*}[t!]
     \centering
     \includegraphics[width=\linewidth]{overview.pdf}
     \caption{Top: An illustration of how users book listings on Airbnb, encompassing two stages: evaluation and comparison. In the evaluation stage, users quickly browse through the available listings to identify those that pique their interest. During the comparison stage, they compare a small selection of candidate listings across various dimensions to make a final decision. Bottom: We model this comparison shopping behavior using pointwise initial ranker and set-wise reranker for the above two stages, respectively. These two networks are co-trained to predict users' booking probabilities.}\label{overview}
 \end{figure*}
{\vspace{-2pt}
\section{Related Work} \label{related_work}
}
 Learning to rank (LTR) has been a popular research topic for decades, with one important area of study focused on evolving loss functions. The pointwise loss predicts action probability (e.g. pCTR, pCVR) for each item independently\cite{McMahan13}. The pair-wise loss looks at two items each time, and converts ranking to a binary classification problem : whether item A is better than item B. \cite{Burges05,burges2010}. The list-wise loss considers the whole list of items and try to approximate the optimal order~\cite{cao07}.

As demonstrated, researchers have progressively advanced from pointwise to pair-wise and list-wise loss functions, continually refining these methods to more effectively optimize ranking metrics such as NDCG. A key insight is that the more explicitly the relative order of items is modeled, the better the alignment between the loss function and ranking metrics. However, despite these advancements in loss function research, the learned scoring functions remain pointwise, meaning that during inference, the model can only compute relevance scores independently. On another note, it is well established that users' choices are influenced by both item position (positional bias~\cite{thorsten17}) and ranking context (contextual bias~\cite{Shashank24,honglei21}). Attention mechanisms have been widely adopted to address contextual bias~\cite{honglei21,Zhi22} and, more generally, to encode ranking context. For instance, self-attention networks have been employed to model ranking context or cross-document interactions~\cite{Qingyao19,Liang20,Rama20,Changhua19}. In this approach, the learned contextual embeddings, combined with each item's own features, are fed into the scoring function, making the resulting relevance score context-dependent.

The application of self-attention networks to encode ranking context, as demonstrated in \cite{Qingyao19,Liang20,Rama20,Changhua19}, has advanced learning-to-rank approaches to the next level, resulting in setwise ranking models. However, these efforts have primarily focused on developing standalone rankers or re-rankers aimed solely at improving ranking metrics. None have addressed the comparison shopping problem, which holds the potential to not only enhance business metrics but also improve user experience.

\section{Proposed framework}\label{framework}

\subsection{Problem Statement} \label{problem_statement}

Given a search query  \(q_i\) , there are \(n\) matched items returned from retrieval stage. The objective of learning to rank is to assign a score \(s_j\) to \(j^{th}\) item  \(x_{i,j}\)  such that those items could be ranked in descending order of scores. In ML terms,  the training example is represented as \(\{q_i, X_i, Y_i\}\), where \(X_i=\{x_{i,j}\}_{j=1}^n\), with \textit{\(x_{i,j}\in R^{m\times1}\)} denoting the feature vector.  \(Y_i = \{{y_{i,j}}\}_{j=1}^n\) where \textit{\(y_{i,j} = \{0,1\}\)} represents the label. In traditional learning-to-rank setup, the goal is to train a model to minimize the loss function \(Loss(f) = \frac{1}{N}\sum_{i}l(f(q_i,X_i), Y_i)\). Here, \(f(q_i,X_i)\) is an univariate scoring function that depends solely on individual item. To model users' comparison shopping behavior, we maintain a similar setup, but modify both the scoring function (model) and the approach used for training and serving the models.

\subsection{Introduction to Comparison Shopping} \label{subsection:introduction}
Before diving into the modeling of comparison shopping, it is important to first understand how users engage in comparison shopping. Our logs indicate that users generally follow these steps when making their final purchase decisions:

 \begin{itemize}[leftmargin=*]
     \item Initial search and evaluation (pointwise Thinking)
     \begin{itemize}
        \item At this stage, users initiate a search query and are presented with a list of items retrieved by the search engine.
        \item Users typically evaluate each of the top-ranked items independently, considering key factors such as price, ratings, and descriptions, etc
        \item Users may click on individual items to obtain detailed information for subsequent comparisons.
     \end{itemize}
     \item Comparative Evaluation (Set-wise Thinking)
     \begin{itemize}
        \item Once a set of potential items has been identified, users enter the comparison stage, where they evaluate the clicked items relative to one another.
        \item Factors like price differences, feature variations, and user reviews are weighed against each other.
        \item Moreover, the initial impression formed during the pointwise evaluation stage plays a significant role, as users continue to factor in their first impressions when making comparisons.
        \item Users typically go back and forth between items, comparing them on various dimensions to determine the best value for their needs.
     \end{itemize}
     \item Decision Making
     \begin{itemize}
        \item After a few rounds of comparisons, users make a final decision based on the information gathered.
        \item This decision could be influenced by specific preferences (e.g., budget, features) and may take place immediately or after additional rounds of research over multiple days.
    \end{itemize}
 \end{itemize}

 Early research efforts \cite{Burges05} \cite{burges2010} \cite{cao07} \cite{McMahan13} primarily focused on modeling users' pointwise thinking behavior, irrespective of the loss function employed, as the scoring functions remain univariate at inference time. More recent studies \cite{Qingyao19} \cite{Liang20} \cite{Rama20} \cite{Changhua19} have incorporated attention mechanisms to capture item interactions, which can be seen as an attempt to model users' set-wise thinking. Although attention-based methods often employ a re-ranker built on top of a ranker that learns pointwise behavior\cite{Weiwen22}, these components are trained independently. While this strategy mimics certain aspects of comparison shopping, it treats the processes as separate rather than integrated, ultimately leading to sub-optimal outcomes.The rationale behind this is that, during the comparison shopping stage, users continue to apply reasoning and information gained through pointwise thinking. As a result, decoupling the re-ranker from the initial ranker would create a substantial disconnect between the model's behavior and the actual decision-making process of users. \par
 In the following section, we will demonstrate this more rigorously through mathematical formalization and introduce our Learning-to-Comparison-Shop (LTCS) framework. Figure \ref{overview} offers an overview: the top part illustrates how an Airbnb user engages in comparison shopping to make a booking decision, while the bottom part depicts how our proposed LTCS framework models this behavior.
{\vspace{-3pt}
\subsection{Learning to Comparison-Shop Framework} \label{subsection:math}
}
As demonstrated by \cite{Weiwen22}, re-rankers that leverage self-attention mechanisms to encode ranking context have seen widespread adoption in multi-stage ranking systems. In this approach, an initial ranking model is trained to estimate the purchase probability \(P( E_{i,j} | q_i, x_{i,j})\) for each item \(x_{i,j}\) retrieved in response to a query \(q_i\), here \(E_{i,j}\) denotes the observed instance of a user purchasing the j-th item. Subsequently, a re-ranker is independently trained to predict the purchase probability \(P(E_{i,j} | q_i, x_{i,j}, C_{i, j, k})\) for each item \(x_{i,j}\) among the top-K items returned by the initial ranking model, with \(C_{i, j, k}\) representing the context of \(x_{i,j}\) relative to the surrounding items in the top-K items. \par

To enhance this framework, we first demonstrate that:
\begin{equation}
\label{eq:statement}
P(E_{i,j} | q_i, x_{i,j}, C_{i, j, k}) \propto P( E_{i,j} | q_i, x_{i,j})P(E_{i,j} | q_i, C_{i, j, k})  
\end{equation}

Here,  \(P( E_{i,j} | q_i, x_{i,j})\) denotes the ranking score produced by the initial ranker, \(P(E_{i,j} | q_i, x_{i,j}, C_{i, j, k})\) represents the ranking score generated by the re-ranker, and \(P(E_{i,j} | q_i, C_{i, j, k})\) models the contextual ranking bias. 

According to Bayesian rule, we could rewrite right side of (1) as 
\begin{equation}
\label{eq:bayesian}
P( E_{i,j} | q_i, x_{i,j})P(E_{i,j} | q_i, C_{i, j, k})
=\frac{P(E_{i,j}, x_{i,j}|q_i)P(E_{i,j}, C_{i, j, k}|q_i)}{P(x_{i,j} | q_i)P(C_{i, j, k} | q_i)}
\end{equation}

Assuming that the items \(\{x_{i,j}\}\) are independent of each other under the condition of query \(q_i\) can be a reasonable approximation in certain contexts. Retrieved items often share similar features due to the query \(q_i\), which may introduce dependencies among these features. However, when modeling users’ comparison-shopping behavior, these common features may not significantly influence the decision-making process. Instead, it is the uncorrelated or unique features that make an item stand out to users. Therefore, assuming independence among items with respect to these distinctive features could be justified, as it emphasizes the attributes most influential in differentiating items during comparison shopping. \par

Under this assumption, we have:
\begin{equation}
P(C_{i, j, k} | q_i)=\prod_{l=1, l\neq j}^{k} P(x_{i,l} | q_i)
\end{equation}

\begin{equation}
\begin{split}
\label{eq:context_q}
&P(x_{i,j} | q_i)P(C_{i, j, k} | q_i) \\
&=P(x_{i,j} | q_i)\prod_{l=1,l\neq j}^{k}P(x_{i,l}|q_i)=P(x_{i,j}, C_{i, j, k} | q_i)
\end{split}
\end{equation}

Similarly, assuming items \(x_{i,j}\) are independent of each other under condition of query \(q_i\) and purchase event \(E_{i,j}\), we have
\begin{equation}
\label{eq:context_qe}
P(x_{i,j} | E_{i,j}, q_i)P(C_{i, j, k} | E_{i,j},q_i)=P(x_{i,j}, C_{i, j, k} | E_{i,j},q_i)
\end{equation}
Thus, applying Bayes’ rule and incorporating Eq\eqref{eq:context_q} and Eq\eqref{eq:context_qe}, Eq \eqref{eq:bayesian} could be rewritten as
\begin{equation}
\begin{split}
\label{eq:proof}
&P( E_{i,j} | q_i, x_{i,j})P(E_{i,j} | q_i, C_{i, j, k}) \\
&=\frac{P(E_{i,j}, x_{i,j}|q_i)P(E_{i,j}, C_{i, j, k}|q_i)}{P(x_{i,j}|q_i)P(C_{i, j, k}|q_i)} \\
&=\frac{P(x_{i,j}|E_{i,j},q_i)P(E_{i,j}|q_i)P(C_{i, j, k}|E_{i,j},q_i)P(E_{i,j}|q_i)}{P(x_{i,j}, C_{i, j, k}|q_i)} \\
&=\frac{P(x_{i,j}, C_{i, j, k}|E_{i,j},q_i)P(E_{i,j}|q_i)P(E_{i,j}|q_i)}{P(x_{i,j}, C_{i, j, k}|q_i)} \\
&= \frac{P(x_{i,j}, C_{i, j, k}, E_{i,j}|q_i)P(E_{i,j}|q_i)}{P(x_{i,j},C_{i, j, k}|q_i)} \\
&= P(E_{i,j} | q_i, x_{i,j}, C_{i, j, k}) P(E_{i,j}|q_i)
\end{split}
\end{equation}
In Eq \eqref{eq:proof}, \(P(E_{i,j}|q_i)\) is a prior which could be considered as constant, so we could prove Eq\eqref{eq:statement} \[P(E_{i,j} | q_i, x_{i,j}, C_{i, j, k}) \propto P( E_{i,j} | q_i, x_{i,j})P(E_{i,j} | q_i, C_{i, j, k})  \]

From Eq\eqref{eq:statement}, we could infer 1) the re-ranker’s score is highly correlated with that of the initial ranker, and 2) the re-ranker’s learning objective inherently encompasses the initial ranker’s objective. These two points are consistent with our intuition and observations in \ref{subsection:introduction}, where we argue that users’ set-wise thinking is not independent of pointwise thinking; rather, they are integrated components of the decision-making process. These insights suggest a clear direction for improving the modeling of users' comparison behavior: co-training the initial ranker and the set-wise re-ranker. This approach offers several potential benefits:
\begin{enumerate}[leftmargin=*]
    \item As demonstrated earlier, the initial ranker and the set-wise re-ranker are highly correlated. Co-training them can allow the models to complement each other, mitigating concerns about negative transfer.
    \item By learning a shared representation of items, reasoning from the pointwise evaluation can be seamlessly transferred to set-wise comparison. Additionally, understanding how users engage in set-wise reasoning enables the initial ranker to more effectively select the top-K items for comparison shopping.
\end{enumerate}

\section{Learning-to-Comparison-Shop (LTCS) System}\label{system} 
\subsection{System Overview}

\begin{figure}[t!]
    \centering
    \includegraphics[width=\linewidth]{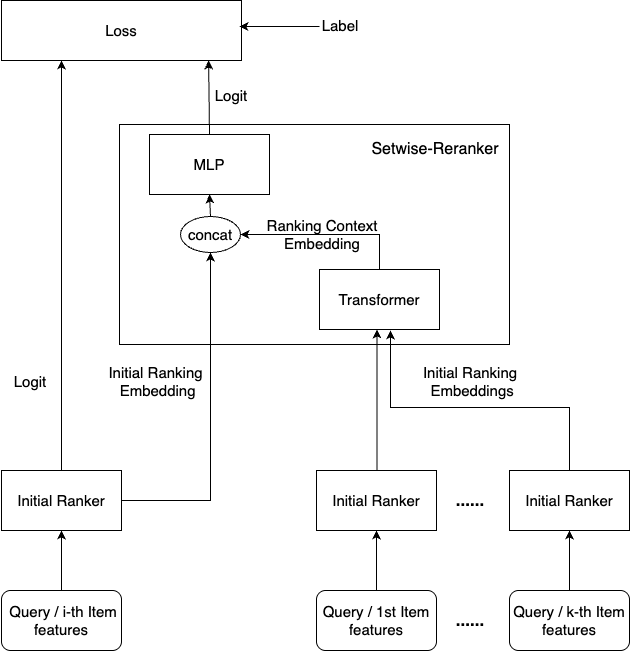}
    \caption{Overall system diagram: For a given candidate item, initial ranker computes its initial ranking logit and also initial ranking embedding based on query and item features. To compute the context embedding, an encoder-only Transformer is employed over top-k items' initial ranking embeddings. Finally the candidate item's initial embedding and context embedding are concatenated and passed through a MLP to generate re-ranker logit. The loss is computed by weighted sum of initial ranker loss and re-ranker loss. }\label{fig_1}
\end{figure}

As outlined in Section \ref{subsection:math}, we propose an innovative ranking system that co-trains the initial ranker alongside a set-wise re-ranker. This system is specifically designed to emulate users’ comparison behavior, with the dual objectives of optimizing key business metrics and improving the overall user experience. The architecture of the model is depicted in Figure \ref{fig_1}:

\begin{enumerate}[leftmargin=*]
    \item The initial ranker is designed as a simple Multi-Layer Perceptron (MLP) network.
    \begin{enumerate}
        \item The input to the network comprises item features and query features.
        \item The network outputs two components: (i) a logit representing the predicted purchase probability, and (ii) an initial ranking embedding. The initial ranking embedding is derived from the final hidden layer of the initial ranker, encapsulating information from each \{query, item\} pair.
        \item The initial rankers in Figure \ref{fig_1} are actually same ranker, all of them share same parameters.
    \end{enumerate}
    \item The set-wise re-ranker is jointly trained with the initial ranker.
    \begin{enumerate}
        \item The initial ranking embeddings of the top-K items (sort by their initial ranker logit scores) are transmitted from the initial ranker to the re-ranker. Within the re-ranker, these embeddings are input into a encoder-only Transformer \cite{devlin2019bert} to produce a context embedding.
        \item Finally, for each item, the re-ranking logit is computed by passing its individual initial ranking embedding and the generated context embedding through the re-ranker’s multilayer perceptron (MLP) network.
    \end{enumerate}
\end{enumerate}

For the purpose of model training, the loss function is defined as follows: Let N denote the number of training examples, where each example corresponds to a single search query. Each training example contains the top K listings returned by the ranking system for that query.

\begin{equation}
\label{eq:initial_loss}
Loss_{initial}=-\sum_{i=1}^{N}\sum_{j=1}^{K}y_{i,j}log(\frac{e^{logit_{initial,i,j}}}{\sum_{k=1}^{K} e^{logit_{initial,i,k}}})
\end{equation}

\begin{equation}
\label{eq:rerank_loss}
Loss_{rerank}=-\sum_{i=1}^{N}\sum_{j=1}^{K}y_{i,j}log(\frac{e^{logit_{rerank,i,j}}}{\sum_{k=1}^{K} e^{logit_{rerank,i,k}}})
\end{equation}

\begin{equation}
\label{eq:loss}
Loss = (1 - \alpha) Loss_{initial} + \alpha Loss_{rerank},
\end{equation}
 
where
\begin{enumerate}
\item \(logit_{initial,i,j}\) is logit output from initial ranker, initial ranker also output item embedding \(embedding_{i,j}\) which is extracted from last hidden layer of initial ranker.
\item \(logit_{rerank,i,j}\) is logit output from reanker.
\end{enumerate}

As shown in Eq\eqref{eq:initial_loss}, Eq\eqref{eq:rerank_loss}, a list-wise loss function \cite{cao07} is applied to both the initial ranker and the re-ranker. The total loss is then obtained by combining these two components through a weighted sum, as described in Eq\eqref{eq:loss}. Additionally, as previously mentioned, the embedding of the \(j-th\) item (\(embedding_{i,j}\)) is extracted from the final hidden layer of the initial ranker. As illustrated in Figure \ref{fig_1}, to compute the re-ranking logit \(logit_{rerank,i,j}\) for each item, its embedding \(embedding_{i,j}\) is concatenated with ranking context embedding \(context\_embedding_{i,j}\) which is derived from an encoder-only Transformer \cite{devlin2019bert} applied to the top-K item embeddings. The concatenated embedding is then passed through the re-ranker MLP for further processing. \par

It is evident that the proposed model structure and training schema are well-aligned with the learning-to-comparison-shop framework outlined in \ref{subsection:math}:
\begin{enumerate}[leftmargin=*]
    \item The initial ranker and re-ranker are co-trained, facilitating the learning of shared item representations.
    \item Knowledge gained from the initial ranker can be transferred to the re-ranker through shared embeddings, while the initial ranker benefits from improved collaboration with the set-wise re-ranker, enabled by these shared embeddings.
\end{enumerate}

\subsection{Model Serving}

In our Learning-to-Comparison-Shop (LTCS) framework, the initial ranker and re-ranker are exported separately from the training graph (Figure \ref{fig_1}) for deployment. To improve serving efficiency, we adopt two key designs:
\begin{itemize}[leftmargin=*]
    \item \textbf{Hierarchical model deployment.} The lightweight, pointwise initial ranker scores all candidates across distributed leaf nodes. The more complex, set-wise re-ranker operates at the master node, refining the top results. This structure enables scalable handling of large candidate pools by expanding leaf nodes.
    \item \textbf{Representation Reuse:} To avoid redundant computation, query-listing representations computed by the initial ranker are reused by the re-ranker via direct transmission from leaf to master nodes, improving efficiency.
\end{itemize}


\section{Experimental Results} \label{exp}

\subsection{Experiment Setup}
The Airbnb ranking system operates as a multi-stage ranking framework. Given a query, the retrieval stage may initially return thousands of listings which are then passed to the initial ranking stage. At this stage, the initial ranker reduces the candidate set, for example, to a few hundred listings. Finally, in the re-ranking stage, few re-rankers further refines this set to optimize for business constraints and objectives. \par
We have successfully deployed the Learning-to-Comparison-Shop (LTCS) framework into our ranking system. However, since the re-ranker employs an encoder-only Transformer to encode the ranking context, its time complexity is \(O(K^2)\), where K represents the number of listings fed into the re-ranker. As a result, K cannot be set too large, as it would substantially increase the search engine’s latency. Therefore, we limit the re-ranking process to the top 40 listings, meaning that only the top 40 results are re-ranked, while the remaining listings retain their original order.

The proposed LTCS models were trained on approximately 360 million examples collected over last one year, using a dataset consisting exclusively of booking labels. The structures of both the initial ranker and re-ranker were defined based on careful hyperparameter tuning to select optimal primary parameters (details provided in Section \ref{hyperparameter}).
\begin{itemize}[leftmargin=*]
    \item As previously mentioned, the initial ranker is implemented as a Multi-Layer Perceptron (MLP) network, with hidden layers consisting of 2048, 1024, 512, 256, and 64 neurons, respectively. The activation function used is Smelu \cite{Rohan22}. The model is fed with de-identified set of hundreds of listing, query, and user features, and is trained to optimize multiple objectives, as outlined in \cite{Jie24}. Under the LTCS framework, the model produces two outputs: a booking probability logit and an initial ranking embedding, which is extracted from the final hidden layer. 
    \item As depicted in Fig \ref{fig_1}, re-ranker's input consists of the initial ranking embeddings of the top-K listings returned from the initial ranking stage. These embeddings are fed into an encoder-only Transformer to generate context embedding. The encoder-only Transformer consists of 30 stacked encoder layers, each comprising a multi-head self-attention sublayer (each layer includes 4 attention heads) and a feedforward sublayer, both equipped with residual connections and layer normalization. The input to the network is the embedding sequence of the top 40 listings generated by the initial ranker. Since, at the comparison-shopping stage, users often browse through all top listings irrespective of their initial ranking order, positional embeddings are not required in this context. For each listing, the initial ranking embedding and the context embedding are concatenated and passed through the re-ranker MLP network with hidden layers consisting of 256, 128, and 64 neurons. The activation function used is also Smelu \cite{Rohan22}. The final order of the top40 listings is determined based on the re-ranker’s logits.
    \item The training loss is computed as the average of the initial ranker loss and the re-ranker loss: \(Loss = 0.5 Loss_{initial} + 0.5Loss_{rerank}\).
\end{itemize}
The model is trained in a distributed manner using 10 A10G GPUs, with a total training time of approximately 20 hours. Prior to deploying LTCS into our ranking system, a three-week online A/B test was conducted, with the control model being our existing multi-objective learning-to-rank model \cite{Jie24}.

\subsection{Baselines}
To the best of our knowledge, LTCS is the first attempt to explicitly address user comparison shopping behavior in an industrial ranking system. As such, there are no direct existing baselines tailored to this specific objective. Therefore, in this study, we compare LTCS against the following strong alternative baselines:
\begin{itemize}[leftmargin=*]
    \item \textbf{attn-DIN} \cite{Rama20} is a state-of-the-art neural re-ranking model\cite{Weiwen22}. It has been shown to outperform several influential neural re-rankers, including SetRank \cite{Liang20}, Deep Listwise Context Model (DLCM) \cite{Qingyao18}, and Groupwise Scoring Functions (GSF) \cite{Qingyao19}. These methods typically treat reranking as an isolated stage applied to a fixed candidate set, optimizing for relevance within that subset.
    \item To ensure a fairer comparison to our LTCS framework—which first uses a pointwise initial ranker(identical to that in LTCS) to filter top-K candidates before applying a setwise re-ranker—we additionally implement a modified version—\textbf{attn-DIN+}—which includes a separately trained initial ranker to simulate a two-stage pipeline. However, unlike LTCS, attn-DIN+ trains the rankers independently, without any shared parameters or coordinated optimization, and does not model the transition from broad item evaluation to focused comparison.
    \item \textbf{MO-LTR} \cite{Jie24}, our current production model, is a multi-objective learning-to-rank framework designed for one-step, end-to-end optimization across multiple business goals. While it has largely replaced traditional re-rankers in our existing ranking stack by learning a unified objective, our evaluation demonstrates that the additional value provided by the LTCS re-ranker cannot be fully captured by MO-LTR alone. 
\end{itemize}

{\subsection{Offline evaluation} \label{subsection:ablation}}

\begin{table}[h]
    \vspace{-8pt}
    \centering
    \begin{tabular}{|l|c|r|}
    \hline
    \textbf{Model} & \textbf{NDCG} & \textbf{NDCG STDEV} \\
    \hline
    attn-DIN & 0.6856 & 0.01 \\
    attn-DIN+ & 0.6918 & 0.008 \\
    MO-LTR & 0.6860 & 0.0004 \\
    LTCS &  \textbf{0.6982} & \textbf{0.0003} \\  
    \hline
    \end{tabular}
    \caption{Offline evaluation results comparing NDCG and its standard deviation across four ranking models}\vspace{-20pt}
    \label{tab:model_comparison}
\end{table}

Table~\ref{tab:model_comparison} presents the offline evaluation results in terms of NDCG and its standard deviation across four ranking models. Among the baselines, attn-DIN and MO-LTR achieve similar NDCG scores (0.6856 and 0.6860 respectively), while attn-DIN+—a two-stage variant with a separately trained initial ranker—improves performance slightly to 0.6918. In contrast, LTCS achieves the highest NDCG score of \textbf{0.6982} (+1.78\% comparing to MO-LTR), outperforming all baselines by a clear margin. Notably, LTCS also exhibits the lowest standard deviation (\textbf{0.0003}), indicating highly stable training dynamics. These results demonstrate that co-training a pointwise initial ranker with a setwise re-ranker not only yields higher ranking effectiveness, but also leads to significantly improved robustness over independently trained or single-stage models.

{
\subsection{Online Evaluation}
}

We evaluate the performance of our new system, Learning-to-Comparison-Shop (LTCS) System with online A/B tests. We conducted a 3-week A/B test, where the control model was multi-objective learning to ranking model (MO-LTR) \cite{Jie24}, and the treatment model was the proposed LTCS system. This A/B test demonstrated a \textbf{+0.6\%} increase in booking conversion rate (CVR) with a p-value of 0.0, marking one of the largest CVR gains for a core ranking model that we have observed in recent years. \par

In addition to optimizing core business metrics, such as booking conversion rate (CVR), LTCS aims to enhance user experience by learning and modeling users’ comparison behaviors during the shopping process. Our A/B tests confirmed improvements in key user engagement metrics. Specifically, the \textit{Average Unique Listing Detail Pages Viewed in the Same Location Before Booking Request} metric decreased by \textbf{0.88\%} with a p-value of 0.001. This finding indicates that, compared to the control group, users in the treatment group viewed significantly fewer listing detail pages before submitting a booking request than users in the control group. Since this metric is often regarded as an indicator of search efficiency, the findings imply that the ranking model improves the comparison-shopping process, reducing the time and effort required to evaluate multiple listings and ultimately enhancing the user experience. \par

 \begin{table} [ht]
     \centering
     \begin{tabular}{|c|c|} \hline 
          booking conversion rate&  +0.6\%\\ \hline
          Listings viewed before booking&           -0.88\%\\ \hline
     \end{tabular}
     \caption{ Online Evaluation Result}
     \vspace{-15pt}
     \label{tab:boosting}
 \end{table}

{
\subsection {Hyper-parameter tuning} \label{hyperparameter}
}
\begin{figure}[t]
  \centering
  \begin{subfigure}[b]{0.5\columnwidth}
    \centering
    \includegraphics[width=1.00\textwidth]{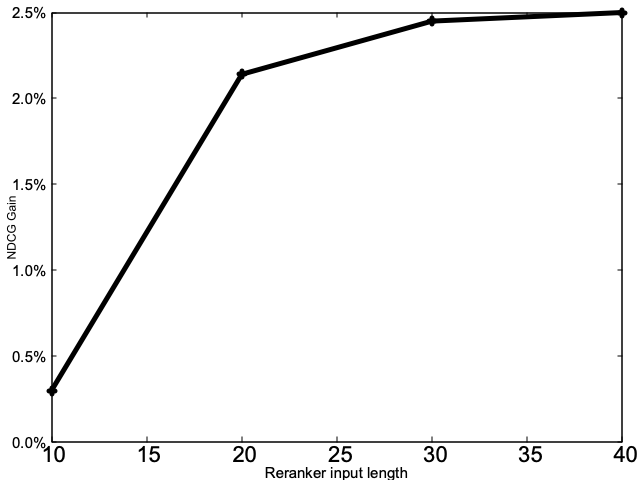}
  \end{subfigure}%
  \begin{subfigure}[b]{0.5\columnwidth}
    \centering
    \includegraphics[width=1.00\textwidth]{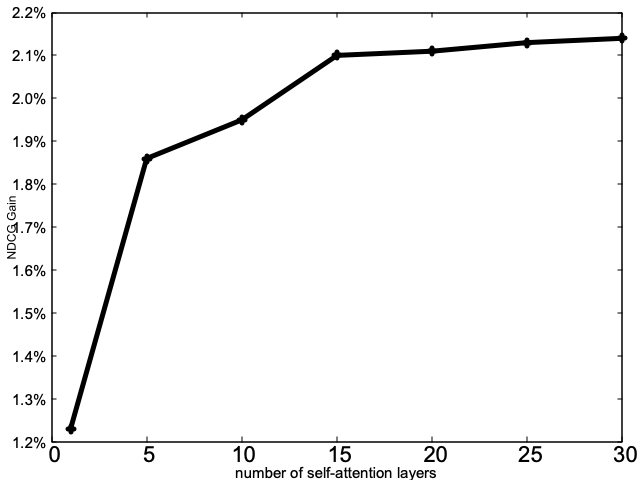}
  \end{subfigure}%
  \vspace{-7pt}
  \caption{Left: Impact of Re-Ranker Input Length on NDCG Gain: Increasing Input Length Leads to Higher NDCG Gains. Right: Effect of encoder-only Transformer Layers on NDCG Gain: Increasing Layers Improves Performance. }
  \label{fig_topk_fig_layers}
  \vspace{-7pt}
\end{figure}

\begin{figure}[t!]
    \centering
    \includegraphics[width=\columnwidth]{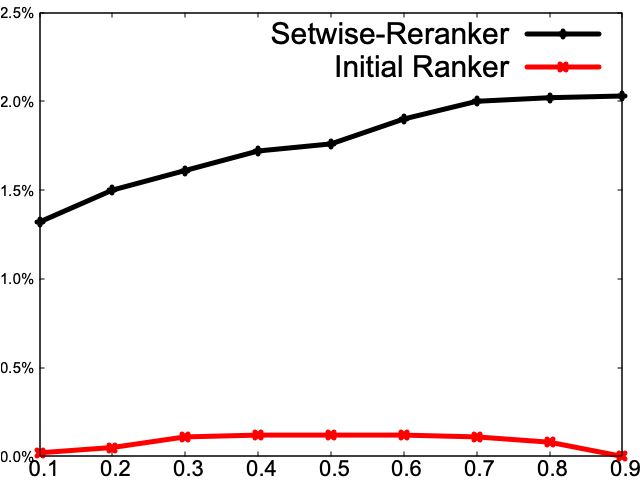}
    \caption{Effect of Re-Ranker Loss Weight on NDCG Gain: Higher Weights Improve Re-Ranker Performance and Benefit the Initial Ranker. The X-axis represents the re-ranker loss weight \(\alpha\), while the Y-axis shows the NDCG gain relative to a ranking system using only the initial ranker.}\vspace{-15pt}
    \label{fig_loss}
\end{figure}

The primary hyperparameters of the LTCS framework are the re-ranker input length and the number of encoder-only Transformer layers. We conducted the following hyperparameter tuning process to select optimal values for these parameters in our deployed model.

\begin{itemize}[leftmargin=*]
    \item Impact of Re-Ranker Input Length: We first examined how varying the re-ranker input length affects model performance. As shown in the left pf Figure~\ref{fig_topk_fig_layers}, increasing the input length from 10 to 40, while keeping the number of Transformer layers fixed at 30, leads to a consistent improvement in NDCG gain (Unless otherwise specified, NDCG gain in this paper refers to the end-to-end NDCG of the full ranking system, encompassing both the initial ranker and the re-ranker). However, beyond an input length of 40, the gain plateaus, indicating diminishing returns. Based on this observation, we selected top-40 as the re-ranking threshold for our deployed model, striking a balance between performance gains and computational efficiency.
    
    \item Impact of Transformer Layers: Next, we investigated the effect of stacking more Transformer layers in the re-ranker. The right-hand side of Figure~\ref{fig_topk_fig_layers} illustrates that increasing the number of layers from 1 to 30, while keeping the input length fixed at 20, results in a steady NDCG improvement. However, beyond 30 layers, the performance gains saturate, suggesting that additional layers provide limited benefit. Consequently, we selected 30 layers for the deployed model to optimize performance while maintaining computational efficiency.
    
    \item Impact of Re-Ranker Loss Weight \(\alpha\): Lastly, we evaluated the role of re-ranker loss weight \(\alpha\) (introduced in Equation \ref{eq:loss}) in model performance. As shown in Figure \ref{fig_loss}, where we consider \(\alpha = 0\) as the baseline (equivalent to a ranking system using only the initial ranker), increasing \(\alpha\) from 0.1 to 0.9 improves the setwise-re-ranker NDCG. Interestingly, co-training also benefits the initial ranker due to the correlation between the two tasks. The initial ranker achieves its peak gain at \(\alpha = 0.5\), but further increasing \(\alpha\) causes the gain to decline. At \(\alpha = 0.9\), the initial ranker no longer shows any improvement, even though the setwise re-ranker reaches its peak. This analysis highlights the trade-off in co-training: further increasing \(\alpha\) enhances setwise re-ranker performance but reduces the initial ranker’s effectiveness as the model prioritizes setwise learning. Based on Figure \ref{fig_loss}, \(\alpha = 0.7\) appears optimal in balancing the two models. However, in our online A/B test, we found that \(\alpha = 0.5\) performed better when considering other business metrics (This falls outside the scope of this paper and will not be further discussed). Therefore, in our deployed model, we set \(\alpha = 0.5\).
    A notable exception occurs when \(\alpha = 1.0\), where only the setwise re-ranker is trained. In this scenario, we observed a significant NDCG drop and highly unstable training dynamics when the same model was trained multiple times. On average, the NDCG gain is 0\% compared to the baseline \(\alpha = 0\), and the standard deviation of NDCG change increased sharply: 0.01 for \(\alpha = 1\) vs. 0.0004 for \(\alpha < 1\). This further demonstrates an additional benefit of co-training the initial ranker and re-ranker: improved training stability. One possible explanation is that during co-training, the initial ranker loss helps regularize the re-ranker’s updates, and vice versa, leading to a more stable optimization process.\cite{Hongxin20}
\end{itemize}

These studies demonstrate that a larger context window and a deeper encoder-only Transformer enable the model to more effectively learn comparison shopping behaviors. This finding aligns with similar observations in the deep learning community and supports our intuition: by reviewing more listings and engaging in deeper cognitive processing, users are likely to make more informed choices or better trade-offs between price and quality. However, it is important to note that users may lose patience with an overly extensive comparison process, the model does not face such limitations.

{\vspace{-7pt}
\section{Conclusion}}

In this study, we introduce the Learning-To-Comparison-Shop (LTCS) system, practiced in Airbnb to model comparison shopping behaviors. First, we ground our design in observations from Airbnb, emphasizing the importance of user initial preferences in comparison shopping. Following this, we detail the efficient implementation of LTCS within a production environment. Through comprehensive experiments and analysis of hyperparameters, we demonstrate that LTCS significantly enhances both ranking and core business metrics in online A/B testing.

\section{GenAI Usage Disclosure}
The authors confirm that no AI-assisted tools were utilized in the writing or preparation of this manuscript.

\bibliographystyle{ACM-Reference-Format}
\balance
\bibliography{sample-base}

@String{Computer = "{IEEE} Computer" }

@InProceedings{Weiwen22,
  author        = {Weiwen Liu and Yuanjia Xi and Jiarui Qin and Fei Sun etc},
  title         = "Neural Re-ranking in Multi-stage Recommender Systems: A Review",
  booktitle     = "Proceedings of the Thirty-First International Joint Conference on Artificial Intelligence (IJCAI-22)",
  year          = {2022},
  pages         = "5512--5520"
}

@InProceedings{Rama20,
  author        = {Rama Kumar Pasumarthi and Honglei Zhuang and Xuanhui Wang etc},
  title         = "Permutation Equivariant Document Interaction Network for Neural Learning-to-Rank",
  booktitle     = "ICTIR '20: Proceedings of the 2020 ACM SIGIR on International Conference on Theory of Information Retrieval",
  year          = {2020},
  pages         = "145--148"
}

@InProceedings{Liang20,
  author        = {Liang Pang and Jun Xu and Qingyao Ai etc},
  title         = "SetRank: Learning a Permutation-Invariant Ranking Model for Information Retrieval",
  booktitle     = "SIGIR '20: Proceedings of the 43rd International ACM SIGIR Conference on Research and Development in Information Retrieval",
  publisher     = "ACM",
  address       = "New York, NY",
  year          = {2020},
  pages         = "499--508"
}

@InProceedings{Qingyao19,
  author        = {Qingyao Ai and Xuanhui Wang and Sebastian Bruch etc},
  title         = "Learning Groupwise Multivariate Scoring Functions Using Deep Neural Networks",
  booktitle     = "ICTIR '19: Proceedings of the 2019 ACM SIGIR International Conference on Theory of Information Retrieval",
  publisher     = "ACM",
  address       = "New York, NY",
  year          = {2019},
  pages         = "85--92"
}

@InProceedings{Changhua19,
  author        = {Changhua Pei and Yi Zhang and Yongfeng Zhang etc},
  title         = "Personalized re-ranking for recommendation",
  booktitle     = "RecSys '19: Proceedings of the 13th ACM Conference on Recommender Systems",
  publisher     = "ACM",
  address       = "New York, NY",
  year          = {2019},
  pages         = "3–11"
}

@InProceedings{Burges05,
  author        = {C. Burges and T. Shaked and E. Renshaw and A. Lazier and M. Deeds and N. Hamilton and G. Hullender},
  title         = "Learning to rank using gradient descent",
  booktitle     = "ICML ’05: Proceedings of the 22nd International Conference on Machine learning",
  year          = {2005},
  pages         = "89--96"
}

@techreport{burges2010,
author = {Burges, Chris J.C.},
title = {From RankNet to LambdaRank to LambdaMART: An Overview},
year = {2010},
month = {June},
number = {MSR-TR-2010-82},
}

@InProceedings{cao07,
  author        = { Z. Cao and T. Qin and T.-Y. Liu and M.-F. Tsai and and H. Li },
  title         = "Learning to rank: from pairwise approach to listwise approach",
  booktitle     = "ICML ’07: Proceedings of the 24th International Conference on Machine learning",
  year          = {2007},
  pages         = "129--136"
}

@InProceedings{McMahan13,
  author        = {H. Brendan McMahan and Gary Holt and D.Sculley and Michael Young etc},
  title         = "Ad Click Prediction: a View from the Trenches",
  booktitle     = "KDD '13: Proceedings of the 19th ACM SIGKDD international conference on Knowledge discovery and data mining",
  publisher     = "ACM",
  address       = "New York, NY",
  year          = {2013},
  pages         = "1222--1230"
}

@InProceedings{thorsten17,
  author        = { Thorsten Jocahims and Adith Swaminathan and Tobias Schnable },
  title         = "Unbiased Learning-to-Rank with Biased Feedback",
  booktitle     = "WSDM '17: Proceedings of the Tenth ACM International Conference on Web Search and Data Mining",
  year          = {2017},
  pages         = "781--789"
}

@InProceedings{Shashank24,
  author        = { Shashank Gupta and Philipp Hager and Jin Huang etc },
  title         = "Unbiased Learning to Rank: On Recent Advances and Practical Applications",
  booktitle     = "WSDM '24: Proceedings of the 17th ACM International Conference on Web Search and Data Mining",
  year          = {2024},
  pages         = "1118--1121"
}

@InProceedings{honglei21,
  author        = { Honglei Zhuang and Zhen Qin and Xuanhui Wang etc },
  title         = "Cross-Positional Attention for Debiasing Clicks",
  booktitle     = "WWW '21: Proceedings of the Web Conference 2021",
  year          = {2021},
  pages         = "788--797"
}

@InProceedings{Zhi22,
  author        = { Zhi Zheng and Zhaopeng Qiu and Tong Xu etc },
  title         = "CBR: Context Bias aware Recommendation for Debiasing User Modeling and Click Prediction",
  booktitle     = "WWW '22: Proceedings of the ACM Web Conference 2022",
  year          = {2022},
  pages         = "2268--2276"
}

@InProceedings{Rohan22,
  author        = { Rohan Anil and Sandra Gadanho and Da Huang and Nijith Jacob etc },
  title         = " On the Factory Floor: ML Engineering for Industrial-Scale Ads Recommendation Models",
  booktitle     = "Recsys 2022 Workshop on Online Recommender Systems and User Modeling",
  year          = {2022}
}

@InProceedings{Jie24,
  author        = { Jie Tang and Huiji Gao and Liwei He and Sanjeev Katariya },
  title         = "Multi-objective Learning to Rank by Model Distillation",
  booktitle     = "KDD '24: Proceedings of the 30th ACM SIGKDD Conference on Knowledge Discovery and Data Mining",
  year          = {2024},
}

@InProceedings{Hongxin20,
  author        = { Hongxin Wei and Lei Feng and Xiangyu Chen and Bo An },
  title         = "Combating Noisy Labels by Agreement: A Joint Training Method with Co-Regularization",
  booktitle     = "2020 IEEE/CVF Conference on Computer Vision and Pattern Recognition (CVPR)",
  year          = {2020},
}

@InProceedings{Qingyao18,
  author        = {  Qingyao Ai and Keping Bi and Jiafeng Guo and W Bruce Croft },
  title         = "Learning a deep listwise context model for ranking refinement",
  booktitle     = "SIGIR",
  year          = {2018},
}

@inproceedings{devlin2019bert,
  title={BERT: Pre-training of Deep Bidirectional Transformers for Language Understanding},
  author={Jacob Devlin, Ming-Wei Chang, Kenton Lee and Kristina Toutanova},
  booktitle={NAACL-HLT},
  year={2019}
}

\end{document}